\documentclass[12pt]{article}
\usepackage{latexsym}
\usepackage{epsfig,graphics}

\newcommand{\be}{\begin{equation}}
\newcommand{\ee}{\end{equation}}

\def\lsi{\raise0.3ex\hbox{$<$\kern-0.75em\raise-1.1ex\hbox{$\sim$}}}
\def\gsi{\raise0.3ex\hbox{$>$\kern-0.75em\raise-1.1ex\hbox{$\sim$}}}

\begin{document}

\begin{titlepage}

\null\vspace{-1.0cm}

\begin{tabbing}
\` Oxford OUTP-00-56P \\
\end{tabbing}

\begin{centering} 
%\vfill

\null\vspace{1.5cm}

{\large \bf  The ${\bf k = 2}$ string tension in four dimensional \\
SU(N) gauge theories}

\vspace{1.5cm}
B. Lucini and M. Teper

\vspace{0.9cm}
{\it Theoretical Physics, University of Oxford, 1 Keble Road, \\
Oxford OX1 3NP, UK\\}

\vspace{2.25cm}
{\bf Abstract.}
\end{centering}

\noindent
We calculate the $k = 2$ string tension in SU(4)
and SU(5) gauge theories in 3+1 dimensions, and compare
it to the $k = 1$ fundamental string tension.
We find, from the continuum extrapolation of 
our lattice calculations, that 
$\sigma_{\mathit{k=2}}/ \sigma_f = 1.40 \pm 0.08$
in the SU(4) gauge theory, and that 
$\sigma_{\mathit{k=2}}/ \sigma_f = 1.56 \pm 0.10$
in SU(5). 
We remark upon the way this might constrain the dynamics 
of confinement and the intriguing implications it might have
for the mass spectrum of SU($N$) gauge theories.
We also note that these results agree closely
with the MQCD-inspired conjecture that the SU($N$) string 
tension varies as $\sigma_{k} \propto \sin(\pi k/N)$.

\vfill

\end{titlepage}

\section{Introduction}
\label{sec_intro}

It is well known that confinement in four dimensional SU($N$) gauge 
theories leads to a linear potential between static test charges in 
the fundamental representation, so that there is a non-zero 
fundamental string tension $\sigma_f \not= 0$. For SU($N\geq 4$) 
there is, however, the possibility of new stable strings which 
join test charges in representations higher than the fundamental.
One can conveniently label these by the way the test charge
transforms under a gauge transformation, $z$, that belongs
to the centre, $Z_N$, of the group. If the charge acquires a
factor $z^k$ we shall refer to the string as having
$N$-ality $k$, and the corresponding string tension
will be denoted by $\sigma_k$. So the fundamental string has
$N$-ality $k=1$ and we shall refer to it as $\sigma_f$ or
$\sigma_{k=1}$, interchangeably.

Because gluons transform trivially under the centre, such a 
$k$-string will not be screened down to a $k^\prime$-string,
if $k^\prime \not= k$. If, however,  $z^k = z^{k^\prime} \ 
\forall z$ then the $k$-string can transform into a
$k^\prime$-string and so one trivially has that
$\sigma_k=\sigma_{k^\prime}$. For example, in the case
of SU(3) a $k=2$ string joining a diquark source to a distant 
anti-diquark source will be simply the usual fundamental 
$k=1$ string over almost all of the separation. 
(In SU(3) three fundamental strings can form a colour singlet.) 
We have to go to at least SU(4)  in order to 
have the possibility of a genuinely different $k=2$ string,
and to SU(6) for a $k=3$ string. 

One can form a $k$-string by simply using static test charges
that consist of $k$ fundamental charges. So the most trivial
possibility is that a `$k$-string' is nothing more than
$k$ separate fundamental strings joining the two test charges. 
In this case we will have $\sigma_k = k\sigma_f$. So for
example, if $\sigma_{k=2} \, {\not <} \, 2\sigma_f$ then the
$k=2$ string is just like two $k=1$ strings and we
have nothing new. There is however a non-trivial possibility.
If $\sigma_{k=2} < 2\sigma_f$ then the  $k=2$ string will
be stable and will constitute a new kind of string. 
We may regard it as a bound state of two  $k=1$ strings. 
Similar comments apply to higher values of $k$.

As emphasised in
\cite{strassler},
one reason why the lattice calculation of $\sigma_k$ 
in SU($N$) gauge theories is particularly interesting is 
that it tests the conjecture 
\cite{strassler}
that the ratio $\sigma_k/\sigma_f$ has a universal
value in `QCD-like' theories:
\be
{{\sigma_k} \over {\sigma_f}} = 
{{\sin \frac{k\pi}{N}}\over{\sin \frac{\pi}{N}}}.
\label{eqn_sigk}
\ee
This conjecture is based on calculations in brane (M-)theory
\cite{strassler}.
The resulting variants of QCD go under the generic name of
MQCD. One finds that in MQCD with ${\cal N} = 1$ 
supersymmetry
\cite{witten,shifman},
the string tensions of $k$-strings satisfy eqn(\ref{eqn_sigk}).
The same is true for ${\cal N} = 2$, for ${\cal N} = 2$ softly
broken to ${\cal N} = 1$ and for non-supersymmetric MQCD
\cite{strassler}.
This is so even though the actual values of $\sigma_k$
differ in these theories. Hence the conjecture 
\cite{strassler}
that the ratio of string tensions is `universal' in
QCD-like theories and takes the value in eqn(\ref{eqn_sigk}). 

Another reason for being interested in the value of
$\sigma_k$ is that the existence of $k$-strings may
have a striking impact on the mass spectrum of SU($N$)
gauge theories in the large-$N$ limit
\cite{largeN}.
A natural model for the glueball states that make
up the spectrum of pure gauge theories is as quantised
closed strings of fundamental flux. Such a picture is 
most compelling for highly excited states which are
large compared to the intrinsic width of the flux tube, 
but may be a reasonably accurate representation for
the low-lying spectrum as well. A simple and explicit model 
based on this idea was suggested some time ago 
\cite{IP}
and comparisons 
\cite{tmmt93,rjmt97,rjmt00}
with the detailed SU($N$) lattice mass spectra available
in 2+1 dimensions
\cite{mt98}
show that it works reasonably well. As emphasised in
\cite{rjmt00}
one can build towers of glueball states on any stable
$k$-string and these will be identical up to a rescaling of the
overall mass scale by a factor of $\sqrt{\sigma_k/\sigma_f}$
(if one neglects the finite width of the flux tube).
This argument neglects mixing and decay, but this 
becomes appropriate as $N\to\infty$. Such a mass spectrum
would provide remarkable evidence for the underlying
string structure of the glueball states. All this does
however depend on the existence of stable $k$-strings
and the details of the SU($N$) spectrum will depend on the
actual values of the corresponding string tensions.

Finally we remark that knowing the value of ${\sigma_k/\sigma_f}$
will constrain the details of any proposed confinement mechanism.

In this paper we will present some results on the $k=2$
string tension and will compare it to the fundamental
($k=1$) tension. We do this for both SU(4) and SU(5),
although our calculations will be more precise
in the former case. There exist previous calculations 
\cite{winoh}
of the lattice SU(4) $k=2$ string tension which show that
$\sigma_k < 2\sigma_f$. However these calculations were
performed at quite a high temperature (although still in the 
confining phase) and were not accurate enough to make a 
useful statement about the continuum limit. We 
intend to perform calculations that are accurate enough,
in the continuum limit, to provide a significant
test of the MQCD conjecture in eqn(\ref{eqn_sigk}).
These calculations are part of our ongoing study 
\cite{blmtN}
of the physical properties of D=3+1 SU($N$) gauge theories,
and a similar calculation is in progress elsewhere
\cite{lddevN}.
\section{Calculations}
\label{sec_calc}

To calculate $\sigma_k$ we first construct a $k$-string
that closes upon itself through a spatial boundary (so 
no explicit static test charges are needed). We then 
calculate the correlation function of two such
$k$-strings as a function of their (Euclidean) time
separation, using standard Monte Carlo simulation 
methods. At sufficiently large separation $t$, the
correlation function will fall as a simple exponential
$\propto \exp(-m_k t)$ where $m_k$ is the mass of the
lightest $k$-string state. Linear confinement means that
this mass will be linear in the length of the string, which 
we denote by $l \equiv aL$ where $a$ is the lattice spacing and 
$L$ is the lattice spatial size in lattice units. In addition 
to this linear dependence the first correction is universal
\cite{Luscherstring}
and we assume that the universality class is that of a simple
bosonic string. (There is evidence from many previous 
SU(2) and SU(3) lattice
calculations that points to this.) Using the string 
correction appropriate to our closed periodic string
\cite{Polystring}
we therefore have
\be
a m_k = a^2 \sigma_k L - \frac{\pi}{3 L} + \ldots
\label{eqn_poly}
\ee
where everything has been expressed in lattice units.
So once we have calculated the mass $a m_k$ we can use
eqn(\ref{eqn_poly}) to extract a value for $a^2\sigma_k$.

In practice a calculation using the simplest 
lattice $k$-string operator is inefficient 
because the overlap onto the lightest string
state is small and so one has to go to large values of 
$t$ before the contribution of excited states has died away;
and there the signal disappears into the statistical noise.
There are standard methods
\cite{blocking}
for curing this problem, using 
blocked (smeared) link operators and variational techniques.
This is described in detail in, for example, 
\cite{mt98}
for the case of a $k=1$ string. Let ${\mathrm{Tr}} \, l_1$ be such a
string, so that $l_1$ is the string before one closes it by 
taking the trace. Then we can form trial $k=2$ strings
from $({\mathrm{Tr}} \,  l_1)^2$  and ${\mathrm{Tr}} \, (l_1 l_1)$. 
In the variational
calculation we allow arbitrary linear combinations of these
in the search for the lightest state. (In practice we
expect this to belong to the appropriate representation
of SU($N$).) Details of this, and of our choice of
Monte Carlo algorithm,
will be given elsewhere
\cite{blmtN}.

Our SU(4) calculations are summarised in Table~\ref{table_datsu4}.
They span a range of couplings that corresponds 
roughly, to the range $\beta \in [5.7,6.0]$ in the case of SU(3)
\cite{MTrev98}.
This is well beyond the bulk phase transition due to the
non-trivial phase structure in the fundamental-adjoint
coupling plane
\cite{mtoldN}.
This phase structure will be more complex as $N$ increases,
since more couplings, appropriate to representations
other than just the fundamental and adjoint, become relevant,
but its effects are expected to remain confined to the region 
between weak and strong coupling. (The phase structure is 
believed to be due to the differing sensitivity of
plaquettes in different representations to the presence of 
$Z_k$ monopoles of various $k$, and there are more possibilities
as $N$ increases.) 

The volumes used in these calculations have been chosen
so that if we were working in SU(3) on the same volumes
(when expressed in physical units e.g. using 
$l_{\sigma} \equiv 1/\surd\sigma_f$)
then eqn(\ref{eqn_poly}) would hold very accurately.
This comparison assumes of course that all these gauge 
theories are similar -- which would be true if they
were all quite close to SU($\infty$). However 
we should not assume that this is so at this stage and
it would be reassuring to have some explicit 
evidence for the assumption of linear confinement, 
and the universal Luscher correction, for
SU(4) and SU(5).
Accordingly we show in Fig.\ref{fig_linear2} how
the flux loop mass varies with its length 
in the SU(4) lattice gauge theory,
for $\beta=10.70$. We see some explicit evidence
that for $L\geq 10$ eqn(\ref{eqn_poly}) is indeed
valid, for both $k=1$ and $k=2$ flux strings. 
For smaller $L$ the higher order corrections to
this relation are significant, especially so for
the $k=2$ string whose string tension appears to  
approach the fundamental string tension as its length
shortens. We have performed a separate calculation
(although only for $k=1$ strings) at the smaller
lattice spacing corresponding to $\beta=10.9$,
and again, as we see in Fig.\ref{fig_linear},
the relation in eqn(\ref{eqn_poly}) holds well
for the volumes used in Table~\ref{table_datsu4}.

\begin	{figure}[p]
\begin	{center}
\leavevmode
% GNUPLOT: LaTeX picture
\setlength{\unitlength}{0.240900pt}
\ifx\plotpoint\undefined\newsavebox{\plotpoint}\fi
\sbox{\plotpoint}{\rule[-0.200pt]{0.400pt}{0.400pt}}%
\begin{picture}(1500,900)(0,0)
\font\gnuplot=cmr10 at 12pt
\gnuplot
\sbox{\plotpoint}{\rule[-0.200pt]{0.400pt}{0.400pt}}%
\put(120.0,31.0){\rule[-0.200pt]{4.818pt}{0.400pt}}
\put(108,31){\makebox(0,0)[r]{{$0$}}}
\put(1436.0,31.0){\rule[-0.200pt]{4.818pt}{0.400pt}}
\put(120.0,109.0){\rule[-0.200pt]{4.818pt}{0.400pt}}
\put(108,109){\makebox(0,0)[r]{{$0.2$}}}
\put(1436.0,109.0){\rule[-0.200pt]{4.818pt}{0.400pt}}
\put(120.0,188.0){\rule[-0.200pt]{4.818pt}{0.400pt}}
\put(108,188){\makebox(0,0)[r]{{$0.4$}}}
\put(1436.0,188.0){\rule[-0.200pt]{4.818pt}{0.400pt}}
\put(120.0,266.0){\rule[-0.200pt]{4.818pt}{0.400pt}}
\put(108,266){\makebox(0,0)[r]{{$0.6$}}}
\put(1436.0,266.0){\rule[-0.200pt]{4.818pt}{0.400pt}}
\put(120.0,344.0){\rule[-0.200pt]{4.818pt}{0.400pt}}
\put(108,344){\makebox(0,0)[r]{{$0.8$}}}
\put(1436.0,344.0){\rule[-0.200pt]{4.818pt}{0.400pt}}
\put(120.0,423.0){\rule[-0.200pt]{4.818pt}{0.400pt}}
\put(108,423){\makebox(0,0)[r]{{$1$}}}
\put(1436.0,423.0){\rule[-0.200pt]{4.818pt}{0.400pt}}
\put(120.0,501.0){\rule[-0.200pt]{4.818pt}{0.400pt}}
\put(108,501){\makebox(0,0)[r]{{$1.2$}}}
\put(1436.0,501.0){\rule[-0.200pt]{4.818pt}{0.400pt}}
\put(120.0,580.0){\rule[-0.200pt]{4.818pt}{0.400pt}}
\put(108,580){\makebox(0,0)[r]{{$1.4$}}}
\put(1436.0,580.0){\rule[-0.200pt]{4.818pt}{0.400pt}}
\put(120.0,658.0){\rule[-0.200pt]{4.818pt}{0.400pt}}
\put(108,658){\makebox(0,0)[r]{{$1.6$}}}
\put(1436.0,658.0){\rule[-0.200pt]{4.818pt}{0.400pt}}
\put(120.0,736.0){\rule[-0.200pt]{4.818pt}{0.400pt}}
\put(108,736){\makebox(0,0)[r]{{$1.8$}}}
\put(1436.0,736.0){\rule[-0.200pt]{4.818pt}{0.400pt}}
\put(120.0,815.0){\rule[-0.200pt]{4.818pt}{0.400pt}}
\put(108,815){\makebox(0,0)[r]{{$2$}}}
\put(1436.0,815.0){\rule[-0.200pt]{4.818pt}{0.400pt}}
\put(120.0,31.0){\rule[-0.200pt]{0.400pt}{4.818pt}}
\put(120,19){\makebox(0,0){\shortstack{\\ \\ \\ {$0$}}}}
\put(120.0,873.0){\rule[-0.200pt]{0.400pt}{4.818pt}}
\put(502.0,31.0){\rule[-0.200pt]{0.400pt}{4.818pt}}
\put(502,19){\makebox(0,0){\shortstack{\\ \\ \\ {$0.05$}}}}
\put(502.0,873.0){\rule[-0.200pt]{0.400pt}{4.818pt}}
\put(883.0,31.0){\rule[-0.200pt]{0.400pt}{4.818pt}}
\put(883,19){\makebox(0,0){\shortstack{\\ \\ \\ {$0.1$}}}}
\put(883.0,873.0){\rule[-0.200pt]{0.400pt}{4.818pt}}
\put(1265.0,31.0){\rule[-0.200pt]{0.400pt}{4.818pt}}
\put(1265,19){\makebox(0,0){\shortstack{\\ \\ \\ {$0.15$}}}}
\put(1265.0,873.0){\rule[-0.200pt]{0.400pt}{4.818pt}}
\put(120.0,31.0){\rule[-0.200pt]{321.842pt}{0.400pt}}
\put(1456.0,31.0){\rule[-0.200pt]{0.400pt}{207.656pt}}
\put(120.0,893.0){\rule[-0.200pt]{321.842pt}{0.400pt}}
\put(-60,558){\makebox(0,0){{\Large{${{{\sigma_{\scriptstyle{k=2}}}{\rule[-0.1in]{0.1in}{0in}}}\over{\sigma_{\scriptstyle{f}}}}$}}}}
\put(788,-89){\makebox(0,0){{\large{$a^2 \sigma_f$}}}}
\put(120.0,31.0){\rule[-0.200pt]{0.400pt}{207.656pt}}
\put(1180,592){\circle*{12}}
\put(837,571){\circle*{12}}
\put(567,563){\circle*{12}}
\put(432,585){\circle*{12}}
\put(1180.0,578.0){\rule[-0.200pt]{0.400pt}{6.745pt}}
\put(1170.0,578.0){\rule[-0.200pt]{4.818pt}{0.400pt}}
\put(1170.0,606.0){\rule[-0.200pt]{4.818pt}{0.400pt}}
\put(837.0,565.0){\rule[-0.200pt]{0.400pt}{3.132pt}}
\put(827.0,565.0){\rule[-0.200pt]{4.818pt}{0.400pt}}
\put(827.0,578.0){\rule[-0.200pt]{4.818pt}{0.400pt}}
\put(567.0,547.0){\rule[-0.200pt]{0.400pt}{7.950pt}}
\put(557.0,547.0){\rule[-0.200pt]{4.818pt}{0.400pt}}
\put(557.0,580.0){\rule[-0.200pt]{4.818pt}{0.400pt}}
\put(432.0,568.0){\rule[-0.200pt]{0.400pt}{7.950pt}}
\put(422.0,568.0){\rule[-0.200pt]{4.818pt}{0.400pt}}
\put(422.0,601.0){\rule[-0.200pt]{4.818pt}{0.400pt}}
\put(1248,634){\circle{12}}
\put(819,615){\circle{12}}
\put(590,642){\circle{12}}
\put(1248.0,611.0){\rule[-0.200pt]{0.400pt}{11.322pt}}
\put(1238.0,611.0){\rule[-0.200pt]{4.818pt}{0.400pt}}
\put(1238.0,658.0){\rule[-0.200pt]{4.818pt}{0.400pt}}
\put(819.0,587.0){\rule[-0.200pt]{0.400pt}{13.249pt}}
\put(809.0,587.0){\rule[-0.200pt]{4.818pt}{0.400pt}}
\put(809.0,642.0){\rule[-0.200pt]{4.818pt}{0.400pt}}
\put(590.0,623.0){\rule[-0.200pt]{0.400pt}{9.395pt}}
\put(580.0,623.0){\rule[-0.200pt]{4.818pt}{0.400pt}}
\put(580.0,662.0){\rule[-0.200pt]{4.818pt}{0.400pt}}
\sbox{\plotpoint}{\rule[-0.500pt]{1.000pt}{1.000pt}}%
\put(120,582){\usebox{\plotpoint}}
\put(120.00,582.00){\usebox{\plotpoint}}
\put(140.72,581.00){\usebox{\plotpoint}}
\multiput(147,581)(20.756,0.000){0}{\usebox{\plotpoint}}
\put(161.47,581.00){\usebox{\plotpoint}}
\put(182.20,580.37){\usebox{\plotpoint}}
\multiput(187,580)(20.756,0.000){0}{\usebox{\plotpoint}}
\put(202.95,580.00){\usebox{\plotpoint}}
\put(223.70,580.00){\usebox{\plotpoint}}
\multiput(228,580)(20.756,0.000){0}{\usebox{\plotpoint}}
\put(244.45,579.75){\usebox{\plotpoint}}
\put(265.18,579.00){\usebox{\plotpoint}}
\multiput(268,579)(20.756,0.000){0}{\usebox{\plotpoint}}
\put(285.93,579.00){\usebox{\plotpoint}}
\put(306.69,579.00){\usebox{\plotpoint}}
\multiput(309,579)(20.694,-1.592){0}{\usebox{\plotpoint}}
\put(327.40,578.00){\usebox{\plotpoint}}
\put(348.16,578.00){\usebox{\plotpoint}}
\multiput(349,578)(20.756,0.000){0}{\usebox{\plotpoint}}
\put(368.92,578.00){\usebox{\plotpoint}}
\put(389.64,577.03){\usebox{\plotpoint}}
\multiput(390,577)(20.756,0.000){0}{\usebox{\plotpoint}}
\put(410.39,577.00){\usebox{\plotpoint}}
\multiput(417,577)(20.756,0.000){0}{\usebox{\plotpoint}}
\put(431.15,577.00){\usebox{\plotpoint}}
\put(451.88,576.39){\usebox{\plotpoint}}
\multiput(457,576)(20.756,0.000){0}{\usebox{\plotpoint}}
\put(472.62,576.00){\usebox{\plotpoint}}
\put(493.37,576.00){\usebox{\plotpoint}}
\multiput(498,576)(20.756,0.000){0}{\usebox{\plotpoint}}
\put(514.12,575.78){\usebox{\plotpoint}}
\put(534.85,575.00){\usebox{\plotpoint}}
\multiput(538,575)(20.756,0.000){0}{\usebox{\plotpoint}}
\put(555.60,575.00){\usebox{\plotpoint}}
\put(576.33,574.19){\usebox{\plotpoint}}
\multiput(579,574)(20.756,0.000){0}{\usebox{\plotpoint}}
\put(597.08,574.00){\usebox{\plotpoint}}
\put(617.84,574.00){\usebox{\plotpoint}}
\multiput(619,574)(20.756,0.000){0}{\usebox{\plotpoint}}
\put(638.57,573.57){\usebox{\plotpoint}}
\put(659.31,573.00){\usebox{\plotpoint}}
\multiput(660,573)(20.756,0.000){0}{\usebox{\plotpoint}}
\put(680.06,573.00){\usebox{\plotpoint}}
\multiput(687,573)(20.756,0.000){0}{\usebox{\plotpoint}}
\put(700.82,572.94){\usebox{\plotpoint}}
\put(721.54,572.00){\usebox{\plotpoint}}
\multiput(727,572)(20.756,0.000){0}{\usebox{\plotpoint}}
\put(742.29,572.00){\usebox{\plotpoint}}
\put(763.05,572.00){\usebox{\plotpoint}}
\multiput(768,572)(20.694,-1.592){0}{\usebox{\plotpoint}}
\put(783.77,571.00){\usebox{\plotpoint}}
\put(804.52,571.00){\usebox{\plotpoint}}
\multiput(808,571)(20.756,0.000){0}{\usebox{\plotpoint}}
\put(825.28,571.00){\usebox{\plotpoint}}
\put(846.01,570.21){\usebox{\plotpoint}}
\multiput(849,570)(20.756,0.000){0}{\usebox{\plotpoint}}
\put(866.75,570.00){\usebox{\plotpoint}}
\put(887.51,570.00){\usebox{\plotpoint}}
\multiput(889,570)(20.756,0.000){0}{\usebox{\plotpoint}}
\put(908.25,569.60){\usebox{\plotpoint}}
\put(928.98,569.00){\usebox{\plotpoint}}
\multiput(930,569)(20.756,0.000){0}{\usebox{\plotpoint}}
\put(949.74,569.00){\usebox{\plotpoint}}
\multiput(957,569)(20.756,0.000){0}{\usebox{\plotpoint}}
\put(970.49,568.96){\usebox{\plotpoint}}
\put(991.21,568.00){\usebox{\plotpoint}}
\multiput(997,568)(20.756,0.000){0}{\usebox{\plotpoint}}
\put(1011.97,568.00){\usebox{\plotpoint}}
\put(1032.70,567.38){\usebox{\plotpoint}}
\multiput(1038,567)(20.756,0.000){0}{\usebox{\plotpoint}}
\put(1053.44,567.00){\usebox{\plotpoint}}
\put(1074.20,567.00){\usebox{\plotpoint}}
\multiput(1078,567)(20.756,0.000){0}{\usebox{\plotpoint}}
\put(1094.95,566.77){\usebox{\plotpoint}}
\put(1115.67,566.00){\usebox{\plotpoint}}
\multiput(1119,566)(20.756,0.000){0}{\usebox{\plotpoint}}
\put(1136.43,566.00){\usebox{\plotpoint}}
\put(1157.18,566.00){\usebox{\plotpoint}}
\multiput(1159,566)(20.703,-1.479){0}{\usebox{\plotpoint}}
\put(1177.90,565.00){\usebox{\plotpoint}}
\put(1198.66,565.00){\usebox{\plotpoint}}
\multiput(1200,565)(20.756,0.000){0}{\usebox{\plotpoint}}
\put(1219.41,565.00){\usebox{\plotpoint}}
\multiput(1227,565)(20.694,-1.592){0}{\usebox{\plotpoint}}
\put(1240.13,564.00){\usebox{\plotpoint}}
\put(1260.89,564.00){\usebox{\plotpoint}}
\multiput(1267,564)(20.756,0.000){0}{\usebox{\plotpoint}}
\put(1281.64,564.00){\usebox{\plotpoint}}
\put(1302.38,563.40){\usebox{\plotpoint}}
\multiput(1308,563)(20.756,0.000){0}{\usebox{\plotpoint}}
\put(1323.12,563.00){\usebox{\plotpoint}}
\put(1343.87,563.00){\usebox{\plotpoint}}
\multiput(1348,563)(20.756,0.000){0}{\usebox{\plotpoint}}
\put(1364.62,562.80){\usebox{\plotpoint}}
\put(1385.34,562.00){\usebox{\plotpoint}}
\multiput(1389,562)(20.756,0.000){0}{\usebox{\plotpoint}}
\put(1406.10,562.00){\usebox{\plotpoint}}
\put(1426.82,561.17){\usebox{\plotpoint}}
\multiput(1429,561)(20.756,0.000){0}{\usebox{\plotpoint}}
\put(1447.57,561.00){\usebox{\plotpoint}}
\put(1456,561){\usebox{\plotpoint}}
\put(120,641){\usebox{\plotpoint}}
\put(120.00,641.00){\usebox{\plotpoint}}
\put(140.74,640.45){\usebox{\plotpoint}}
\multiput(147,640)(20.756,0.000){0}{\usebox{\plotpoint}}
\put(161.48,640.00){\usebox{\plotpoint}}
\put(182.23,640.00){\usebox{\plotpoint}}
\multiput(187,640)(20.756,0.000){0}{\usebox{\plotpoint}}
\put(202.99,640.00){\usebox{\plotpoint}}
\put(223.74,640.00){\usebox{\plotpoint}}
\multiput(228,640)(20.694,-1.592){0}{\usebox{\plotpoint}}
\put(244.46,639.00){\usebox{\plotpoint}}
\put(265.21,639.00){\usebox{\plotpoint}}
\multiput(268,639)(20.756,0.000){0}{\usebox{\plotpoint}}
\put(285.97,639.00){\usebox{\plotpoint}}
\put(306.73,639.00){\usebox{\plotpoint}}
\multiput(309,639)(20.756,0.000){0}{\usebox{\plotpoint}}
\put(327.48,639.00){\usebox{\plotpoint}}
\put(348.20,638.06){\usebox{\plotpoint}}
\multiput(349,638)(20.756,0.000){0}{\usebox{\plotpoint}}
\put(368.95,638.00){\usebox{\plotpoint}}
\put(389.71,638.00){\usebox{\plotpoint}}
\multiput(390,638)(20.756,0.000){0}{\usebox{\plotpoint}}
\put(410.46,638.00){\usebox{\plotpoint}}
\multiput(417,638)(20.756,0.000){0}{\usebox{\plotpoint}}
\put(431.22,638.00){\usebox{\plotpoint}}
\put(451.95,637.39){\usebox{\plotpoint}}
\multiput(457,637)(20.756,0.000){0}{\usebox{\plotpoint}}
\put(472.69,637.00){\usebox{\plotpoint}}
\put(493.45,637.00){\usebox{\plotpoint}}
\multiput(498,637)(20.756,0.000){0}{\usebox{\plotpoint}}
\put(514.20,637.00){\usebox{\plotpoint}}
\put(534.96,637.00){\usebox{\plotpoint}}
\multiput(538,637)(20.703,-1.479){0}{\usebox{\plotpoint}}
\put(555.68,636.00){\usebox{\plotpoint}}
\put(576.43,636.00){\usebox{\plotpoint}}
\multiput(579,636)(20.756,0.000){0}{\usebox{\plotpoint}}
\put(597.19,636.00){\usebox{\plotpoint}}
\put(617.95,636.00){\usebox{\plotpoint}}
\multiput(619,636)(20.756,0.000){0}{\usebox{\plotpoint}}
\put(638.70,636.00){\usebox{\plotpoint}}
\put(659.42,635.04){\usebox{\plotpoint}}
\multiput(660,635)(20.756,0.000){0}{\usebox{\plotpoint}}
\put(680.18,635.00){\usebox{\plotpoint}}
\multiput(687,635)(20.756,0.000){0}{\usebox{\plotpoint}}
\put(700.93,635.00){\usebox{\plotpoint}}
\put(721.69,635.00){\usebox{\plotpoint}}
\multiput(727,635)(20.756,0.000){0}{\usebox{\plotpoint}}
\put(742.44,635.00){\usebox{\plotpoint}}
\put(763.17,634.34){\usebox{\plotpoint}}
\multiput(768,634)(20.756,0.000){0}{\usebox{\plotpoint}}
\put(783.92,634.00){\usebox{\plotpoint}}
\put(804.67,634.00){\usebox{\plotpoint}}
\multiput(808,634)(20.756,0.000){0}{\usebox{\plotpoint}}
\put(825.43,634.00){\usebox{\plotpoint}}
\put(846.18,634.00){\usebox{\plotpoint}}
\multiput(849,634)(20.694,-1.592){0}{\usebox{\plotpoint}}
\put(866.90,633.00){\usebox{\plotpoint}}
\put(887.66,633.00){\usebox{\plotpoint}}
\multiput(889,633)(20.756,0.000){0}{\usebox{\plotpoint}}
\put(908.41,633.00){\usebox{\plotpoint}}
\put(929.17,633.00){\usebox{\plotpoint}}
\multiput(930,633)(20.756,0.000){0}{\usebox{\plotpoint}}
\put(949.92,633.00){\usebox{\plotpoint}}
\multiput(957,633)(20.694,-1.592){0}{\usebox{\plotpoint}}
\put(970.64,632.00){\usebox{\plotpoint}}
\put(991.40,632.00){\usebox{\plotpoint}}
\multiput(997,632)(20.756,0.000){0}{\usebox{\plotpoint}}
\put(1012.15,632.00){\usebox{\plotpoint}}
\put(1032.91,632.00){\usebox{\plotpoint}}
\multiput(1038,632)(20.756,0.000){0}{\usebox{\plotpoint}}
\put(1053.66,632.00){\usebox{\plotpoint}}
\put(1074.39,631.28){\usebox{\plotpoint}}
\multiput(1078,631)(20.756,0.000){0}{\usebox{\plotpoint}}
\put(1095.14,631.00){\usebox{\plotpoint}}
\put(1115.89,631.00){\usebox{\plotpoint}}
\multiput(1119,631)(20.756,0.000){0}{\usebox{\plotpoint}}
\put(1136.65,631.00){\usebox{\plotpoint}}
\put(1157.40,631.00){\usebox{\plotpoint}}
\multiput(1159,631)(20.703,-1.479){0}{\usebox{\plotpoint}}
\put(1178.12,630.00){\usebox{\plotpoint}}
\put(1198.88,630.00){\usebox{\plotpoint}}
\multiput(1200,630)(20.756,0.000){0}{\usebox{\plotpoint}}
\put(1219.63,630.00){\usebox{\plotpoint}}
\multiput(1227,630)(20.756,0.000){0}{\usebox{\plotpoint}}
\put(1240.39,630.00){\usebox{\plotpoint}}
\put(1261.14,630.00){\usebox{\plotpoint}}
\multiput(1267,630)(20.703,-1.479){0}{\usebox{\plotpoint}}
\put(1281.86,629.00){\usebox{\plotpoint}}
\put(1302.62,629.00){\usebox{\plotpoint}}
\multiput(1308,629)(20.756,0.000){0}{\usebox{\plotpoint}}
\put(1323.37,629.00){\usebox{\plotpoint}}
\put(1344.13,629.00){\usebox{\plotpoint}}
\multiput(1348,629)(20.756,0.000){0}{\usebox{\plotpoint}}
\put(1364.88,628.78){\usebox{\plotpoint}}
\put(1385.60,628.00){\usebox{\plotpoint}}
\multiput(1389,628)(20.756,0.000){0}{\usebox{\plotpoint}}
\put(1406.36,628.00){\usebox{\plotpoint}}
\put(1427.11,628.00){\usebox{\plotpoint}}
\multiput(1429,628)(20.756,0.000){0}{\usebox{\plotpoint}}
\put(1447.87,628.00){\usebox{\plotpoint}}
\put(1456,628){\usebox{\plotpoint}}
\end{picture}

\end	{center}
\vskip 0.15in
\caption{The ratio of $k=2$ and $k=1$ string tensions in 
our SU(4)($\bullet$) and SU(5) ($\circ$) lattice
calculations plotted as a function of $a^2\sigma_f$.
Extrapolations to the continuum limit, using a leading 
$O(a^2)$ correction, are displayed.}
\label{fig_sig2}
\end 	{figure}
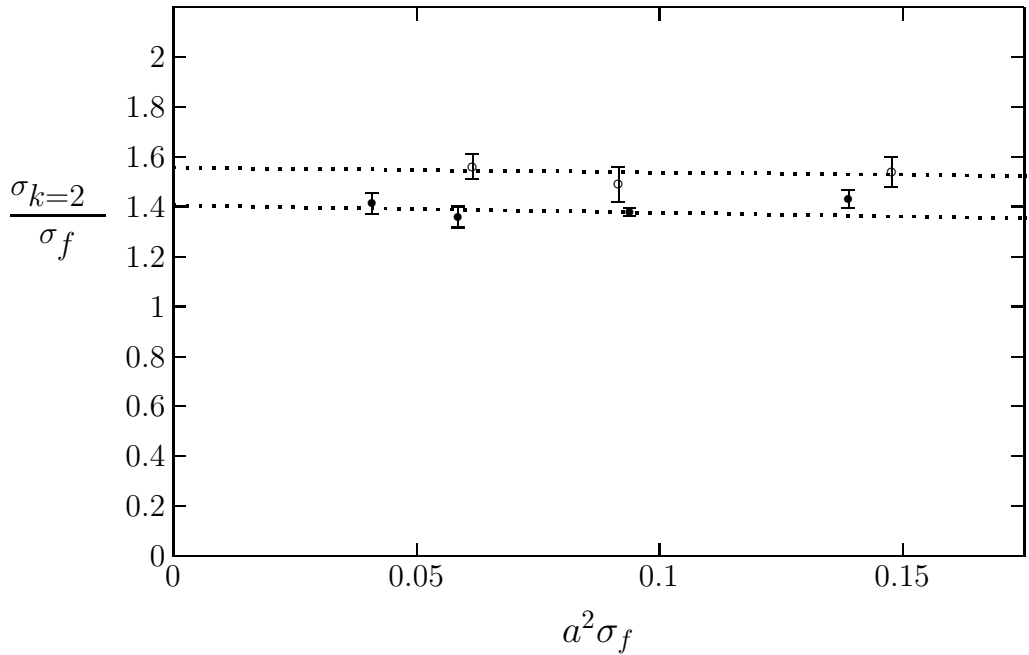

With the standard plaquette action that we are using, the
leading lattice correction to the continuum limit is $O(a^2)$,
and so for sufficiently small $a$ we can extrapolate
to the continuum using
\be
{{\sigma_k(a)} \over {\sigma_f(a)}} =
{{\sigma_k(0)} \over {\sigma_f(0)}} + c a^2\sigma_f .
\label{eqn_extrap}
\ee
In Fig.\ref{fig_sig2} we plot our calculated values of 
${{a^2\sigma_{\mathit{k=2}}}/{a^2\sigma_f}} \equiv
\sigma_{\mathit{k=2}}/ \sigma_f$ 
against the simultaneously calculated value of $a^2\sigma_f$. 
On such a plot the continuum extrapolation in 
eqn(\ref{eqn_extrap}) is a simple straight line.
It is clear that all our points are consistent with lying
on such a line and we show the best fit in the plot.
From this we extract our continuum limit for the
ratio of string tensions:
\be
\lim_{a \to 0}
{{\sigma_{\mathit{k=2}}} \over {\sigma_f}}
=
\left\{ \begin{array}{ll}
1.354 \pm 0.070  & \ \ \ \beta \geq 10.55 \\
1.405 \pm 0.080  & \ \ \ \beta \geq 10.70
\end{array}
\right. \ \ \ \ \ \ \ \ \ : \ {\mathrm{SU(4)}}.
\label{eqn_contsu4}
\ee
Both of the fits in eqn(\ref{eqn_contsu4}) are quite good,
with confidence levels $\sim 40\%$. (The error limits
correspond to confidence levels $\sim 10\%$.)
The $\beta \geq 10.70$
fit has the virtue of avoiding the $\beta = 10.55$ calculation 
which might be biased by the nearby bulk phase transition,
but suffers from a shorter lever arm.

Our SU(5) calculations, which are more limited in scope,
are summarised in Table~\ref{table_datsu5} and 
the results for $\sigma_{\mathit{k=2}} / \sigma_f$
are plotted in Fig.\ref{fig_sig2}.
The continuum extrapolation, using eqn(\ref{eqn_extrap}),
is clearly less constrained than for SU(4) and leads to:
\be
\lim_{a \to 0}
{{\sigma_{\mathit{k=2}}} \over {\sigma_f}} 
=
1.556 \pm 0.100 \ \ \ \ \ \ \ \ \ :  \ {\mathrm{SU(5)}}.
\label{eqn_contsu5}
\ee
In addition we have deliberately tuned the SU(5) couplings 
so that each of the three values of $a$ roughly coincides, 
when expressed in physical units (of $1/\surd\sigma_f$), 
with one of the SU(4) couplings. It is plausible that
the lattice corrections largely cancel in the relative
SU(4) and SU(5) ratios, so that the average 
\be
{  ({\sigma_{\mathit{k=2}}}/{\sigma_f})_{\scriptscriptstyle{SU(5)}} 
\over
({\sigma_{\mathit{k=2}}}/{\sigma_f})_{\scriptscriptstyle{SU(4)}} }
=
1.10 \pm 0.03   \ \ \ \ \ \  \ \ \ : \ \
0.38 \geq a\sqrt{\sigma_f} \geq 0.24 
\label{eqn_rat45}
\ee
should be close to its  continuum value.

\section{Discussion}
\label{sec_disc}

According to the MQCD conjecture in eqn(\ref{eqn_sigk})
we should expect
\be
{{\sigma_{\mathit{k=2}}}\over{\sigma_f}}
=
{{\sin \frac{2\pi}{N}}\over{\sin \frac{\pi}{N}}} 
=
2 \cos \frac{\pi}{N}
=
\left\{ \begin{array}{ll}
0  & \ \ \ {\mathrm{SU(2)}} \\
1  & \ \ \ {\mathrm{SU(3)}} \\
1.41 ... & \ \ \ {\mathrm{SU(4)}} \\
1.61 ... & \ \ \ {\mathrm{SU(5)}}
\end{array}
\right.
\label{eqn_sig2b}
\ee
This is plausible for SU(2), where two fundamental charges
can form an unconfined colour singlet, and for SU(3),
where two fundamental strings can couple to 
a single fundamental (anti-)string. (We say `plausible'
because one could imagine the $N\geq 4$ string bound
states continuing down to unstable, resonant string states
for $N=3$ and $N=2$.)  Our SU(4) results
are consistent with eqn(\ref{eqn_sig2b}), within
fairly small errors. So are our SU(5) results; although
the continuum extrapolation is less accurate.
The ratio in eqn(\ref{eqn_rat45}) is indeed close to the
MQCD prediction of $\simeq 1.14$. All this appears
to provide significant support for the approximate
validity of the MQCD conjecture. 

To gauge how significant this apparent agreement is,
it would be useful to compare with something else.
Since our numerical calculations are on a Euclidean
lattice with a plaquette action, one obvious comparison 
is with the strong coupling prediction. This is
well-known to be 
\be
{{\sigma_{\mathit{k}}}\over{\sigma_f}}
= 
\min \{k,N-k\}
\stackrel{\mathit{k=2}}{=}
\left\{ \begin{array}{ll}
0  & \ \ \ {\mathrm{SU(2)}} \\
1  & \ \ \ {\mathrm{SU(3)}} \\
2  & \ \ \ {\mathrm{SU(N\geq 4)}}
\end{array}
\right.
\label{eqn_euclid}
\ee
which starts out well but is quite inconsistent with
our SU(4) and SU(5) calculations. A more intriguing
comparison, emphasised in
\cite{strassler},
is with the Hamiltonian strong coupling result
\be
{{\sigma_{\mathit{k}}}\over{\sigma_f}}
= 
{{k(N-k)}\over{N-1}}
\stackrel{\mathit{k=2}}{=}
\left\{ \begin{array}{ll}
0  & \ \ \ {\mathrm{SU(2)}} \\
1  & \ \ \ {\mathrm{SU(3)}} \\
1.3{\bar 3}  & \ \ \ {\mathrm{SU(4)}} \\
1.50   & \ \ \ {\mathrm{SU(5)}}
\end{array}
\right.
\label{eqn_hamilton}
\ee
This is clearly compatible with our numerical results.
Of course one might question the relevance of 
comparing Euclidean lattice calculations with
the strong coupling limit of the Hamiltonian
lattice approach. After all the strong coupling
limit is not universal and  we could
find almost anything we wanted in some strong
coupling limit or other. (For example, if we were
to construct a lattice action that uses link matrices
in matrix representations other than just the 
fundamental.) On the other hand
the origin of the formula in eqn(\ref{eqn_hamilton})
has a nice simple physical interpretation that
might arise in a more general context, and one might
argue that colour-electric flux is better represented
with time continuous. It is therefore particularly
interesting to note that our errors are
already comparable to the difference between the
MQCD and Hamiltonian strong coupling predictions.
A calculation that is better than ours by,
say, a factor of $O(10)$ would readily discriminate 
between eqn(\ref{eqn_euclid}) and eqn(\ref{eqn_hamilton}).
Such a calculation is quite feasible and we intend to
carry it out in the near future.

In any case, we have shown that bound {$k=2$}
strings states do exist for SU($N\geq 4$) and that 
the binding is strong for SU(4) and SU(5). This will have
implications for the confinement dynamics. For example
\footnote{this argument arose in a discussion with
C. Korthals Altes},
in the picture where confinement in SU($N$) gauge theories
is driven by the condensation of centre vortices 
\cite{thooftvor}
the value of $\sigma_{\mathit{k=2}}/\sigma_f$ provides
information on the relative density of $k=2$ and
$k=1$ vortices. If one neglects all correlations
between vortices, then the calculation of how
vortices disorder Wilson loops involves only
a simple Poisson distribution
\cite{poissonvor}
and one readily finds that the number of $k=2$ vortices 
piercing unit area (of the minimal surface spanning a
Wilson loop) is related to the number of $k=1$ vortices by
\be
{{N^v_{\mathit{k=2}}}\over{N^v_{\mathit{k=1}}}}
=
{{\sigma_{\mathit{k=1}}}\over{\sigma_{\mathit{k=2}}}}
-
\frac{1}{2}
\label{eqn_sigvor}
\ee
in the case of SU(4). So using our calculated string 
tensions we obtain a constraint on the relative density of
doubly charged centre vortices condensed in the SU(4)
vacuum: $N^v_2/N^v_1 \simeq 0.25(5)$. It is interesting to 
note that at high temperatures one can calculate the
dual 't Hooft string tension in perturbation theory,
and one finds
\cite{ckavor}
a formula that happens to be identical to the Hamiltonian
strong coupling result given in eqn(\ref{eqn_hamilton}).

The fact that the $k=2$ string tension is much less than 
twice the fundamental string tension, for $N=4$ and $N=5$, 
suggests the possibility that the mass spectra of these 
gauge theories might contain prominent states based upon 
closed loops of such {$k=2$} strings, that are additional to 
any states based on closed loops of fundamental flux.
It would clearly be interesting to perform lattice 
calculations of SU($N$) gauge theories that are designed
to explore this intriguing possibility.

%\vspace{1.5cm}
%
%
%\noindent { {\bf Acknowledgements:}}
\section*{Acknowledgments}
We are grateful to M. Strassler for emphasising to one of 
us (MT), some time ago, the interest of such calculations and
for useful communications as this paper was being written. 
We also thank C. Korthals Altes,  L. Del Debbio and Alex Kovner
for interesting discussions. Our calculations
were carried out on Alpha Compaq workstations in Oxford
Theoretical Physics, funded by PPARC and EPSRC grants.
One of us (BL) thanks PPARC for a postdoctoral fellowship.

\vfill \eject

\begin{table}
\begin{center}
\begin{tabular}{|c|c|c||c|c|}\hline
\multicolumn{5}{|c|}{SU(4)} \\ \hline
$\beta$ & lattice & MC sweeps & $a\surd\sigma_f$ & 
$\sigma_{k=2}/\sigma_f$ \\ \hline
10.55 & $8^4$  & $2\times 10^5$ & 0.3726(30) & 1.432(35) \\
10.70 & $10^4$ & $10^5$ & 0.3070(12) & 1.380(19) \\
10.70 & $12^4$ & $10^5$ & 0.3058(12) & 1.377(35) \\
10.90 & $12^4$ & $10^5$ & 0.2421(17) & 1.358(42) \\ 
11.10 & $16^4$ & $9.6\times 10^4$ & 0.2022(11) & 1.413(42) \\ \hline
\end{tabular}
\caption{\label{table_datsu4}
The ratio of $k=2$ and $k=1$ string tensions in SU(4) from lattice
calculations with various lattice spacings (given in units of the
fundamental string tension).}
\end{center}
\end{table}

\begin{table}
\begin{center}
\begin{tabular}{|c|c|c||c|c|}\hline
\multicolumn{5}{|c|}{SU(5)} \\ \hline
$\beta$ & lattice & MC sweeps & $a\surd\sigma_f$ & 
$\sigma_{k=2}/\sigma_f$ \\ \hline
16.755 & $8^4$  & $10^5$             & 0.3845(20) & 1.54(6) \\
16.975 & $10^4$ & $1.6\times 10^5$   & 0.3027(21) & 1.49(7) \\ 
17.270 & $12^4$ & $8\times 10^4$     & 0.2480(11) & 1.56(5) \\ 
\hline
\end{tabular}
\caption{\label{table_datsu5}
The ratio of $k=2$ and $k=1$ string tensions in SU(5) for lattice
calculations with various lattice spacings (given in units of the
fundamental string tension).}
\end{center}
\end{table}

\newpage

\begin	{figure}[p]
\begin	{center}
\leavevmode
% GNUPLOT: LaTeX picture
\setlength{\unitlength}{0.240900pt}
\ifx\plotpoint\undefined\newsavebox{\plotpoint}\fi
\sbox{\plotpoint}{\rule[-0.200pt]{0.400pt}{0.400pt}}%
\begin{picture}(1500,900)(0,0)
\font\gnuplot=cmr10 at 12pt
\gnuplot
\sbox{\plotpoint}{\rule[-0.200pt]{0.400pt}{0.400pt}}%
\put(120.0,31.0){\rule[-0.200pt]{4.818pt}{0.400pt}}
\put(108,31){\makebox(0,0)[r]{{$0$}}}
\put(1436.0,31.0){\rule[-0.200pt]{4.818pt}{0.400pt}}
\put(120.0,127.0){\rule[-0.200pt]{4.818pt}{0.400pt}}
\put(108,127){\makebox(0,0)[r]{{$0.2$}}}
\put(1436.0,127.0){\rule[-0.200pt]{4.818pt}{0.400pt}}
\put(120.0,223.0){\rule[-0.200pt]{4.818pt}{0.400pt}}
\put(108,223){\makebox(0,0)[r]{{$0.4$}}}
\put(1436.0,223.0){\rule[-0.200pt]{4.818pt}{0.400pt}}
\put(120.0,318.0){\rule[-0.200pt]{4.818pt}{0.400pt}}
\put(108,318){\makebox(0,0)[r]{{$0.6$}}}
\put(1436.0,318.0){\rule[-0.200pt]{4.818pt}{0.400pt}}
\put(120.0,414.0){\rule[-0.200pt]{4.818pt}{0.400pt}}
\put(108,414){\makebox(0,0)[r]{{$0.8$}}}
\put(1436.0,414.0){\rule[-0.200pt]{4.818pt}{0.400pt}}
\put(120.0,510.0){\rule[-0.200pt]{4.818pt}{0.400pt}}
\put(108,510){\makebox(0,0)[r]{{$1$}}}
\put(1436.0,510.0){\rule[-0.200pt]{4.818pt}{0.400pt}}
\put(120.0,606.0){\rule[-0.200pt]{4.818pt}{0.400pt}}
\put(108,606){\makebox(0,0)[r]{{$1.2$}}}
\put(1436.0,606.0){\rule[-0.200pt]{4.818pt}{0.400pt}}
\put(120.0,701.0){\rule[-0.200pt]{4.818pt}{0.400pt}}
\put(108,701){\makebox(0,0)[r]{{$1.4$}}}
\put(1436.0,701.0){\rule[-0.200pt]{4.818pt}{0.400pt}}
\put(120.0,797.0){\rule[-0.200pt]{4.818pt}{0.400pt}}
\put(108,797){\makebox(0,0)[r]{{$1.6$}}}
\put(1436.0,797.0){\rule[-0.200pt]{4.818pt}{0.400pt}}
\put(120.0,31.0){\rule[-0.200pt]{0.400pt}{4.818pt}}
\put(120,19){\makebox(0,0){\shortstack{\\ \\ \\ {$0$}}}}
\put(120.0,873.0){\rule[-0.200pt]{0.400pt}{4.818pt}}
\put(454.0,31.0){\rule[-0.200pt]{0.400pt}{4.818pt}}
\put(454,19){\makebox(0,0){\shortstack{\\ \\ \\ {$4$}}}}
\put(454.0,873.0){\rule[-0.200pt]{0.400pt}{4.818pt}}
\put(788.0,31.0){\rule[-0.200pt]{0.400pt}{4.818pt}}
\put(788,19){\makebox(0,0){\shortstack{\\ \\ \\ {$8$}}}}
\put(788.0,873.0){\rule[-0.200pt]{0.400pt}{4.818pt}}
\put(1122.0,31.0){\rule[-0.200pt]{0.400pt}{4.818pt}}
\put(1122,19){\makebox(0,0){\shortstack{\\ \\ \\ {$12$}}}}
\put(1122.0,873.0){\rule[-0.200pt]{0.400pt}{4.818pt}}
\put(1456.0,31.0){\rule[-0.200pt]{0.400pt}{4.818pt}}
\put(1456,19){\makebox(0,0){\shortstack{\\ \\ \\ {$16$}}}}
\put(1456.0,873.0){\rule[-0.200pt]{0.400pt}{4.818pt}}
\put(120.0,31.0){\rule[-0.200pt]{321.842pt}{0.400pt}}
\put(1456.0,31.0){\rule[-0.200pt]{0.400pt}{207.656pt}}
\put(120.0,893.0){\rule[-0.200pt]{321.842pt}{0.400pt}}
\put(-60,558){\makebox(0,0){{\Large{$m_{\scriptscriptstyle{k}}$}}}}
\put(788,-89){\makebox(0,0){{\large{$L$}}}}
\put(120.0,31.0){\rule[-0.200pt]{0.400pt}{207.656pt}}
\put(621,158){\circle*{12}}
\put(788,301){\circle*{12}}
\put(955,432){\circle*{12}}
\put(1122,527){\circle*{12}}
\put(621.0,154.0){\rule[-0.200pt]{0.400pt}{2.168pt}}
\put(611.0,154.0){\rule[-0.200pt]{4.818pt}{0.400pt}}
\put(611.0,163.0){\rule[-0.200pt]{4.818pt}{0.400pt}}
\put(788.0,296.0){\rule[-0.200pt]{0.400pt}{2.168pt}}
\put(778.0,296.0){\rule[-0.200pt]{4.818pt}{0.400pt}}
\put(778.0,305.0){\rule[-0.200pt]{4.818pt}{0.400pt}}
\put(955.0,429.0){\rule[-0.200pt]{0.400pt}{1.686pt}}
\put(945.0,429.0){\rule[-0.200pt]{4.818pt}{0.400pt}}
\put(945.0,436.0){\rule[-0.200pt]{4.818pt}{0.400pt}}
\put(1122.0,522.0){\rule[-0.200pt]{0.400pt}{2.168pt}}
\put(1112.0,522.0){\rule[-0.200pt]{4.818pt}{0.400pt}}
\put(1112.0,531.0){\rule[-0.200pt]{4.818pt}{0.400pt}}
\put(621,184){\circle{12}}
\put(788,399){\circle{12}}
\put(955,603){\circle{12}}
\put(1122,729){\circle{12}}
\put(621.0,178.0){\rule[-0.200pt]{0.400pt}{3.132pt}}
\put(611.0,178.0){\rule[-0.200pt]{4.818pt}{0.400pt}}
\put(611.0,191.0){\rule[-0.200pt]{4.818pt}{0.400pt}}
\put(788.0,387.0){\rule[-0.200pt]{0.400pt}{5.541pt}}
\put(778.0,387.0){\rule[-0.200pt]{4.818pt}{0.400pt}}
\put(778.0,410.0){\rule[-0.200pt]{4.818pt}{0.400pt}}
\put(955.0,596.0){\rule[-0.200pt]{0.400pt}{3.613pt}}
\put(945.0,596.0){\rule[-0.200pt]{4.818pt}{0.400pt}}
\put(945.0,611.0){\rule[-0.200pt]{4.818pt}{0.400pt}}
\put(1122.0,712.0){\rule[-0.200pt]{0.400pt}{8.431pt}}
\put(1112.0,712.0){\rule[-0.200pt]{4.818pt}{0.400pt}}
\put(1112.0,747.0){\rule[-0.200pt]{4.818pt}{0.400pt}}
\sbox{\plotpoint}{\rule[-0.500pt]{1.000pt}{1.000pt}}%
\put(399.00,31.00){\usebox{\plotpoint}}
\put(413.15,46.15){\usebox{\plotpoint}}
\put(427.83,60.83){\usebox{\plotpoint}}
\put(442.96,75.03){\usebox{\plotpoint}}
\multiput(444,76)(14.676,14.676){0}{\usebox{\plotpoint}}
\put(457.72,89.62){\usebox{\plotpoint}}
\put(473.49,103.11){\usebox{\plotpoint}}
\put(489.31,116.55){\usebox{\plotpoint}}
\put(505.10,130.01){\usebox{\plotpoint}}
\put(521.25,143.05){\usebox{\plotpoint}}
\put(537.67,155.74){\usebox{\plotpoint}}
\multiput(538,156)(16.320,12.823){0}{\usebox{\plotpoint}}
\put(554.01,168.54){\usebox{\plotpoint}}
\put(570.60,181.00){\usebox{\plotpoint}}
\put(587.27,193.36){\usebox{\plotpoint}}
\put(604.44,205.00){\usebox{\plotpoint}}
\multiput(606,206)(16.451,12.655){0}{\usebox{\plotpoint}}
\put(621.10,217.35){\usebox{\plotpoint}}
\put(638.24,229.03){\usebox{\plotpoint}}
\put(655.23,240.93){\usebox{\plotpoint}}
\put(672.40,252.58){\usebox{\plotpoint}}
\multiput(673,253)(17.459,11.224){0}{\usebox{\plotpoint}}
\put(689.78,263.92){\usebox{\plotpoint}}
\put(707.00,275.50){\usebox{\plotpoint}}
\put(724.23,287.08){\usebox{\plotpoint}}
\multiput(727,289)(17.459,11.224){0}{\usebox{\plotpoint}}
\put(741.63,298.39){\usebox{\plotpoint}}
\put(759.24,309.37){\usebox{\plotpoint}}
\put(776.50,320.89){\usebox{\plotpoint}}
\put(794.27,331.58){\usebox{\plotpoint}}
\multiput(795,332)(17.065,11.814){0}{\usebox{\plotpoint}}
\put(811.56,343.04){\usebox{\plotpoint}}
\put(829.18,353.97){\usebox{\plotpoint}}
\put(846.88,364.79){\usebox{\plotpoint}}
\multiput(849,366)(17.677,10.878){0}{\usebox{\plotpoint}}
\put(864.65,375.51){\usebox{\plotpoint}}
\put(882.31,386.37){\usebox{\plotpoint}}
\put(899.96,397.26){\usebox{\plotpoint}}
\multiput(903,399)(17.677,10.878){0}{\usebox{\plotpoint}}
\put(917.73,407.99){\usebox{\plotpoint}}
\put(935.64,418.47){\usebox{\plotpoint}}
\put(953.52,429.01){\usebox{\plotpoint}}
\multiput(957,431)(17.065,11.814){0}{\usebox{\plotpoint}}
\put(970.81,440.46){\usebox{\plotpoint}}
\put(988.74,450.91){\usebox{\plotpoint}}
\put(1006.60,461.48){\usebox{\plotpoint}}
\multiput(1011,464)(17.677,10.878){0}{\usebox{\plotpoint}}
\put(1024.37,472.19){\usebox{\plotpoint}}
\put(1042.70,481.89){\usebox{\plotpoint}}
\put(1060.56,492.46){\usebox{\plotpoint}}
\multiput(1065,495)(17.677,10.878){0}{\usebox{\plotpoint}}
\put(1078.33,503.19){\usebox{\plotpoint}}
\put(1096.27,513.63){\usebox{\plotpoint}}
\put(1114.12,524.21){\usebox{\plotpoint}}
\put(1131.89,534.93){\usebox{\plotpoint}}
\multiput(1132,535)(18.564,9.282){0}{\usebox{\plotpoint}}
\put(1150.23,544.61){\usebox{\plotpoint}}
\put(1168.08,555.19){\usebox{\plotpoint}}
\put(1185.86,565.91){\usebox{\plotpoint}}
\multiput(1186,566)(18.564,9.282){0}{\usebox{\plotpoint}}
\put(1204.20,575.59){\usebox{\plotpoint}}
\put(1222.05,586.17){\usebox{\plotpoint}}
\put(1239.82,596.89){\usebox{\plotpoint}}
\multiput(1240,597)(18.564,9.282){0}{\usebox{\plotpoint}}
\put(1258.17,606.56){\usebox{\plotpoint}}
\put(1276.02,617.15){\usebox{\plotpoint}}
\multiput(1281,620)(18.275,9.840){0}{\usebox{\plotpoint}}
\put(1294.22,627.12){\usebox{\plotpoint}}
\put(1312.16,637.56){\usebox{\plotpoint}}
\put(1330.28,647.64){\usebox{\plotpoint}}
\multiput(1335,650)(17.677,10.878){0}{\usebox{\plotpoint}}
\put(1348.18,658.11){\usebox{\plotpoint}}
\put(1366.26,668.30){\usebox{\plotpoint}}
\put(1384.41,678.38){\usebox{\plotpoint}}
\multiput(1389,681)(17.677,10.878){0}{\usebox{\plotpoint}}
\put(1402.18,689.09){\usebox{\plotpoint}}
\put(1420.52,698.78){\usebox{\plotpoint}}
\put(1438.66,708.83){\usebox{\plotpoint}}
\multiput(1443,711)(17.677,10.878){0}{\usebox{\plotpoint}}
\put(1456,719){\usebox{\plotpoint}}
\put(357.00,31.00){\usebox{\plotpoint}}
\put(368.86,48.02){\usebox{\plotpoint}}
\put(380.70,65.04){\usebox{\plotpoint}}
\put(393.41,81.46){\usebox{\plotpoint}}
\put(406.16,97.83){\usebox{\plotpoint}}
\put(419.33,113.87){\usebox{\plotpoint}}
\put(432.53,129.89){\usebox{\plotpoint}}
\put(446.18,145.52){\usebox{\plotpoint}}
\put(459.89,161.10){\usebox{\plotpoint}}
\put(474.05,176.28){\usebox{\plotpoint}}
\put(488.33,191.33){\usebox{\plotpoint}}
\put(502.82,206.19){\usebox{\plotpoint}}
\put(517.17,221.17){\usebox{\plotpoint}}
\put(531.85,235.85){\usebox{\plotpoint}}
\put(546.53,250.53){\usebox{\plotpoint}}
\put(561.56,264.83){\usebox{\plotpoint}}
\put(576.78,278.94){\usebox{\plotpoint}}
\put(591.54,293.54){\usebox{\plotpoint}}
\multiput(592,294)(15.759,13.508){0}{\usebox{\plotpoint}}
\put(607.18,307.18){\usebox{\plotpoint}}
\put(622.06,321.62){\usebox{\plotpoint}}
\put(637.67,335.31){\usebox{\plotpoint}}
\put(653.15,349.13){\usebox{\plotpoint}}
\put(668.62,362.96){\usebox{\plotpoint}}
\put(684.23,376.63){\usebox{\plotpoint}}
\put(699.57,390.61){\usebox{\plotpoint}}
\multiput(700,391)(15.759,13.508){0}{\usebox{\plotpoint}}
\put(715.32,404.12){\usebox{\plotpoint}}
\put(731.15,417.55){\usebox{\plotpoint}}
\put(746.94,431.02){\usebox{\plotpoint}}
\put(762.73,444.49){\usebox{\plotpoint}}
\put(778.55,457.93){\usebox{\plotpoint}}
\put(794.80,470.84){\usebox{\plotpoint}}
\multiput(795,471)(15.251,14.078){0}{\usebox{\plotpoint}}
\put(810.21,484.73){\usebox{\plotpoint}}
\put(826.39,497.72){\usebox{\plotpoint}}
\put(842.46,510.86){\usebox{\plotpoint}}
\put(858.49,524.03){\usebox{\plotpoint}}
\put(874.71,536.98){\usebox{\plotpoint}}
\multiput(876,538)(15.844,13.407){0}{\usebox{\plotpoint}}
\put(890.64,550.29){\usebox{\plotpoint}}
\put(906.84,563.25){\usebox{\plotpoint}}
\put(922.89,576.41){\usebox{\plotpoint}}
\put(938.94,589.56){\usebox{\plotpoint}}
\put(955.14,602.54){\usebox{\plotpoint}}
\multiput(957,604)(15.844,13.407){0}{\usebox{\plotpoint}}
\put(971.10,615.79){\usebox{\plotpoint}}
\put(987.75,628.17){\usebox{\plotpoint}}
\put(1003.79,641.33){\usebox{\plotpoint}}
\put(1019.84,654.48){\usebox{\plotpoint}}
\put(1036.46,666.90){\usebox{\plotpoint}}
\multiput(1038,668)(15.844,13.407){0}{\usebox{\plotpoint}}
\put(1052.44,680.13){\usebox{\plotpoint}}
\put(1068.65,693.09){\usebox{\plotpoint}}
\put(1084.92,705.95){\usebox{\plotpoint}}
\put(1101.21,718.79){\usebox{\plotpoint}}
\put(1117.85,731.18){\usebox{\plotpoint}}
\multiput(1119,732)(15.844,13.407){0}{\usebox{\plotpoint}}
\put(1133.81,744.43){\usebox{\plotpoint}}
\put(1150.17,757.21){\usebox{\plotpoint}}
\put(1166.56,769.94){\usebox{\plotpoint}}
\put(1182.96,782.66){\usebox{\plotpoint}}
\put(1199.30,795.45){\usebox{\plotpoint}}
\multiput(1200,796)(16.451,12.655){0}{\usebox{\plotpoint}}
\put(1215.73,808.14){\usebox{\plotpoint}}
\put(1232.09,820.91){\usebox{\plotpoint}}
\put(1248.47,833.66){\usebox{\plotpoint}}
\put(1264.88,846.37){\usebox{\plotpoint}}
\multiput(1267,848)(16.320,12.823){0}{\usebox{\plotpoint}}
\put(1281.22,859.17){\usebox{\plotpoint}}
\put(1297.64,871.86){\usebox{\plotpoint}}
\put(1314.01,884.62){\usebox{\plotpoint}}
\multiput(1321,890)(16.604,12.453){0}{\usebox{\plotpoint}}
\put(1325,893){\usebox{\plotpoint}}
\end{picture}

\end	{center}
\vskip 0.15in
\caption{The masses of the fundamental, $\bullet$, and $k=2$,
$\circ$, flux loops versus their length, at $\beta=10.7$ for SU(4).
Shown also is the variation one expects from
eqn(\ref{eqn_poly}).}
\label{fig_linear2}
\end 	{figure}
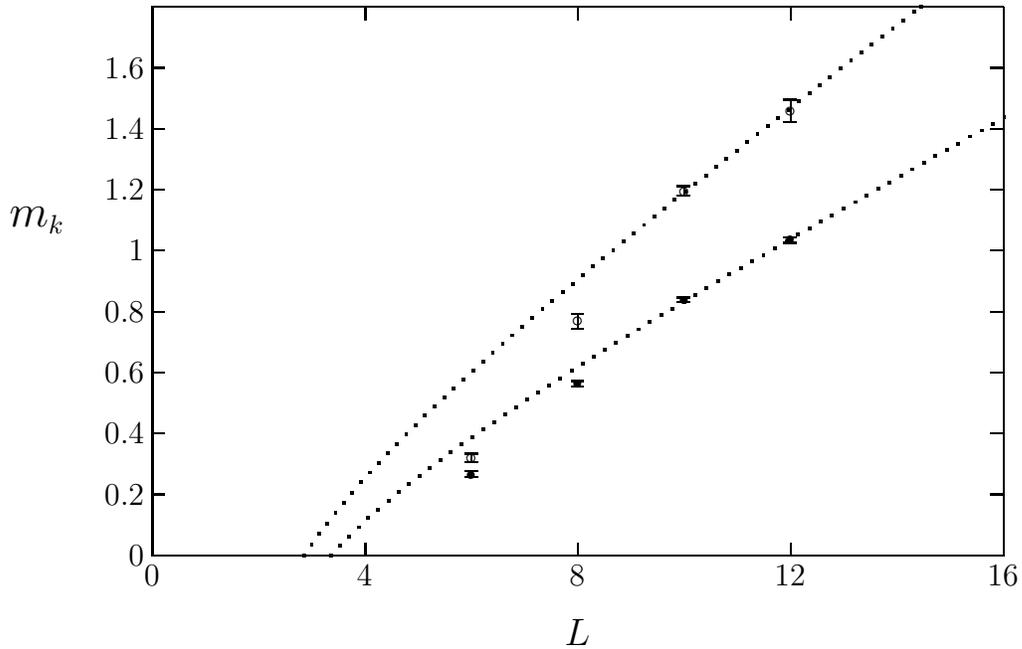

\begin	{figure}[p]
\begin	{center}
\leavevmode
% GNUPLOT: LaTeX picture
\setlength{\unitlength}{0.240900pt}
\ifx\plotpoint\undefined\newsavebox{\plotpoint}\fi
\sbox{\plotpoint}{\rule[-0.200pt]{0.400pt}{0.400pt}}%
\begin{picture}(1500,900)(0,0)
\font\gnuplot=cmr10 at 12pt
\gnuplot
\sbox{\plotpoint}{\rule[-0.200pt]{0.400pt}{0.400pt}}%
\put(120.0,31.0){\rule[-0.200pt]{4.818pt}{0.400pt}}
\put(108,31){\makebox(0,0)[r]{{$0$}}}
\put(1436.0,31.0){\rule[-0.200pt]{4.818pt}{0.400pt}}
\put(120.0,223.0){\rule[-0.200pt]{4.818pt}{0.400pt}}
\put(108,223){\makebox(0,0)[r]{{$0.2$}}}
\put(1436.0,223.0){\rule[-0.200pt]{4.818pt}{0.400pt}}
\put(120.0,414.0){\rule[-0.200pt]{4.818pt}{0.400pt}}
\put(108,414){\makebox(0,0)[r]{{$0.4$}}}
\put(1436.0,414.0){\rule[-0.200pt]{4.818pt}{0.400pt}}
\put(120.0,606.0){\rule[-0.200pt]{4.818pt}{0.400pt}}
\put(108,606){\makebox(0,0)[r]{{$0.6$}}}
\put(1436.0,606.0){\rule[-0.200pt]{4.818pt}{0.400pt}}
\put(120.0,797.0){\rule[-0.200pt]{4.818pt}{0.400pt}}
\put(108,797){\makebox(0,0)[r]{{$0.8$}}}
\put(1436.0,797.0){\rule[-0.200pt]{4.818pt}{0.400pt}}
\put(120.0,31.0){\rule[-0.200pt]{0.400pt}{4.818pt}}
\put(120,19){\makebox(0,0){\shortstack{\\ \\ \\ {$4$}}}}
\put(120.0,873.0){\rule[-0.200pt]{0.400pt}{4.818pt}}
\put(565.0,31.0){\rule[-0.200pt]{0.400pt}{4.818pt}}
\put(565,19){\makebox(0,0){\shortstack{\\ \\ \\ {$8$}}}}
\put(565.0,873.0){\rule[-0.200pt]{0.400pt}{4.818pt}}
\put(1011.0,31.0){\rule[-0.200pt]{0.400pt}{4.818pt}}
\put(1011,19){\makebox(0,0){\shortstack{\\ \\ \\ {$12$}}}}
\put(1011.0,873.0){\rule[-0.200pt]{0.400pt}{4.818pt}}
\put(1456.0,31.0){\rule[-0.200pt]{0.400pt}{4.818pt}}
\put(1456,19){\makebox(0,0){\shortstack{\\ \\ \\ {$16$}}}}
\put(1456.0,873.0){\rule[-0.200pt]{0.400pt}{4.818pt}}
\put(120.0,31.0){\rule[-0.200pt]{321.842pt}{0.400pt}}
\put(1456.0,31.0){\rule[-0.200pt]{0.400pt}{207.656pt}}
\put(120.0,893.0){\rule[-0.200pt]{321.842pt}{0.400pt}}
\put(-60,558){\makebox(0,0){{\Large{$m_{\scriptscriptstyle{k=1}}$}}}}
\put(788,-89){\makebox(0,0){{\large{$L$}}}}
\put(120.0,31.0){\rule[-0.200pt]{0.400pt}{207.656pt}}
\put(565,295){\circle*{12}}
\put(788,457){\circle*{12}}
\put(1011,590){\circle*{12}}
\put(565.0,286.0){\rule[-0.200pt]{0.400pt}{4.095pt}}
\put(555.0,286.0){\rule[-0.200pt]{4.818pt}{0.400pt}}
\put(555.0,303.0){\rule[-0.200pt]{4.818pt}{0.400pt}}
\put(788.0,434.0){\rule[-0.200pt]{0.400pt}{11.081pt}}
\put(778.0,434.0){\rule[-0.200pt]{4.818pt}{0.400pt}}
\put(778.0,480.0){\rule[-0.200pt]{4.818pt}{0.400pt}}
\put(1011.0,574.0){\rule[-0.200pt]{0.400pt}{7.709pt}}
\put(1001.0,574.0){\rule[-0.200pt]{4.818pt}{0.400pt}}
\put(1001.0,606.0){\rule[-0.200pt]{4.818pt}{0.400pt}}
\sbox{\plotpoint}{\rule[-0.500pt]{1.000pt}{1.000pt}}%
\put(157.00,31.00){\usebox{\plotpoint}}
\put(171.96,45.25){\usebox{\plotpoint}}
\put(186.78,59.78){\usebox{\plotpoint}}
\multiput(187,60)(15.759,13.508){0}{\usebox{\plotpoint}}
\put(202.47,73.36){\usebox{\plotpoint}}
\put(217.98,87.13){\usebox{\plotpoint}}
\put(234.12,100.18){\usebox{\plotpoint}}
\put(249.92,113.64){\usebox{\plotpoint}}
\put(266.14,126.57){\usebox{\plotpoint}}
\multiput(268,128)(16.320,12.823){0}{\usebox{\plotpoint}}
\put(282.47,139.39){\usebox{\plotpoint}}
\put(298.53,152.52){\usebox{\plotpoint}}
\put(315.25,164.81){\usebox{\plotpoint}}
\put(331.96,177.12){\usebox{\plotpoint}}
\put(348.52,189.63){\usebox{\plotpoint}}
\multiput(349,190)(16.889,12.064){0}{\usebox{\plotpoint}}
\put(365.42,201.67){\usebox{\plotpoint}}
\put(382.42,213.58){\usebox{\plotpoint}}
\put(399.40,225.51){\usebox{\plotpoint}}
\put(416.78,236.86){\usebox{\plotpoint}}
\multiput(417,237)(16.451,12.655){0}{\usebox{\plotpoint}}
\put(433.44,249.21){\usebox{\plotpoint}}
\put(450.75,260.67){\usebox{\plotpoint}}
\put(468.06,272.11){\usebox{\plotpoint}}
\multiput(471,274)(17.677,10.878){0}{\usebox{\plotpoint}}
\put(485.68,283.08){\usebox{\plotpoint}}
\put(503.02,294.48){\usebox{\plotpoint}}
\put(520.60,305.48){\usebox{\plotpoint}}
\put(537.90,316.93){\usebox{\plotpoint}}
\multiput(538,317)(18.021,10.298){0}{\usebox{\plotpoint}}
\put(555.70,327.56){\usebox{\plotpoint}}
\put(573.20,338.69){\usebox{\plotpoint}}
\put(590.99,349.38){\usebox{\plotpoint}}
\multiput(592,350)(17.459,11.224){0}{\usebox{\plotpoint}}
\put(608.49,360.53){\usebox{\plotpoint}}
\put(626.31,371.18){\usebox{\plotpoint}}
\put(644.11,381.84){\usebox{\plotpoint}}
\multiput(646,383)(18.021,10.298){0}{\usebox{\plotpoint}}
\put(662.06,392.27){\usebox{\plotpoint}}
\put(679.86,402.92){\usebox{\plotpoint}}
\put(697.68,413.57){\usebox{\plotpoint}}
\multiput(700,415)(18.021,10.298){0}{\usebox{\plotpoint}}
\put(715.62,424.00){\usebox{\plotpoint}}
\put(733.42,434.67){\usebox{\plotpoint}}
\put(751.24,445.30){\usebox{\plotpoint}}
\multiput(754,447)(18.564,9.282){0}{\usebox{\plotpoint}}
\put(769.59,454.98){\usebox{\plotpoint}}
\put(787.39,465.65){\usebox{\plotpoint}}
\put(805.55,475.68){\usebox{\plotpoint}}
\multiput(808,477)(18.021,10.298){0}{\usebox{\plotpoint}}
\put(823.58,485.97){\usebox{\plotpoint}}
\put(841.57,496.28){\usebox{\plotpoint}}
\put(859.60,506.52){\usebox{\plotpoint}}
\multiput(862,508)(18.564,9.282){0}{\usebox{\plotpoint}}
\put(877.95,516.20){\usebox{\plotpoint}}
\put(895.96,526.48){\usebox{\plotpoint}}
\put(913.97,536.75){\usebox{\plotpoint}}
\multiput(916,538)(18.564,9.282){0}{\usebox{\plotpoint}}
\put(932.32,546.42){\usebox{\plotpoint}}
\put(950.34,556.67){\usebox{\plotpoint}}
\put(968.34,566.98){\usebox{\plotpoint}}
\multiput(970,568)(18.564,9.282){0}{\usebox{\plotpoint}}
\put(986.77,576.49){\usebox{\plotpoint}}
\put(1004.94,586.54){\usebox{\plotpoint}}
\put(1023.13,596.53){\usebox{\plotpoint}}
\multiput(1024,597)(18.564,9.282){0}{\usebox{\plotpoint}}
\put(1041.50,606.15){\usebox{\plotpoint}}
\put(1059.59,616.29){\usebox{\plotpoint}}
\put(1077.95,625.97){\usebox{\plotpoint}}
\multiput(1078,626)(18.564,9.282){0}{\usebox{\plotpoint}}
\put(1096.30,635.64){\usebox{\plotpoint}}
\put(1114.42,645.71){\usebox{\plotpoint}}
\multiput(1119,648)(18.275,9.840){0}{\usebox{\plotpoint}}
\put(1132.78,655.39){\usebox{\plotpoint}}
\put(1151.26,664.83){\usebox{\plotpoint}}
\put(1169.39,674.94){\usebox{\plotpoint}}
\multiput(1173,677)(18.275,9.840){0}{\usebox{\plotpoint}}
\put(1187.64,684.82){\usebox{\plotpoint}}
\put(1206.11,694.29){\usebox{\plotpoint}}
\put(1224.56,703.78){\usebox{\plotpoint}}
\multiput(1227,705)(18.275,9.840){0}{\usebox{\plotpoint}}
\put(1242.92,713.46){\usebox{\plotpoint}}
\put(1261.37,722.97){\usebox{\plotpoint}}
\put(1279.84,732.42){\usebox{\plotpoint}}
\multiput(1281,733)(17.677,10.878){0}{\usebox{\plotpoint}}
\put(1297.75,742.88){\usebox{\plotpoint}}
\put(1316.19,752.41){\usebox{\plotpoint}}
\put(1334.68,761.84){\usebox{\plotpoint}}
\multiput(1335,762)(18.275,9.840){0}{\usebox{\plotpoint}}
\put(1353.04,771.52){\usebox{\plotpoint}}
\put(1371.45,781.09){\usebox{\plotpoint}}
\multiput(1375,783)(18.564,9.282){0}{\usebox{\plotpoint}}
\put(1389.94,790.51){\usebox{\plotpoint}}
\put(1408.32,800.16){\usebox{\plotpoint}}
\put(1426.71,809.77){\usebox{\plotpoint}}
\multiput(1429,811)(18.564,9.282){0}{\usebox{\plotpoint}}
\put(1445.20,819.19){\usebox{\plotpoint}}
\put(1456,825){\usebox{\plotpoint}}
\end{picture}

\end	{center}
\vskip 0.15in
\caption{The mass of the fundamental flux loop 
as a function of its length, at $\beta=10.9$ for SU(4).
Shown also is the variation one expects from
eqn(\ref{eqn_poly}).}
\label{fig_linear}
\end 	{figure}
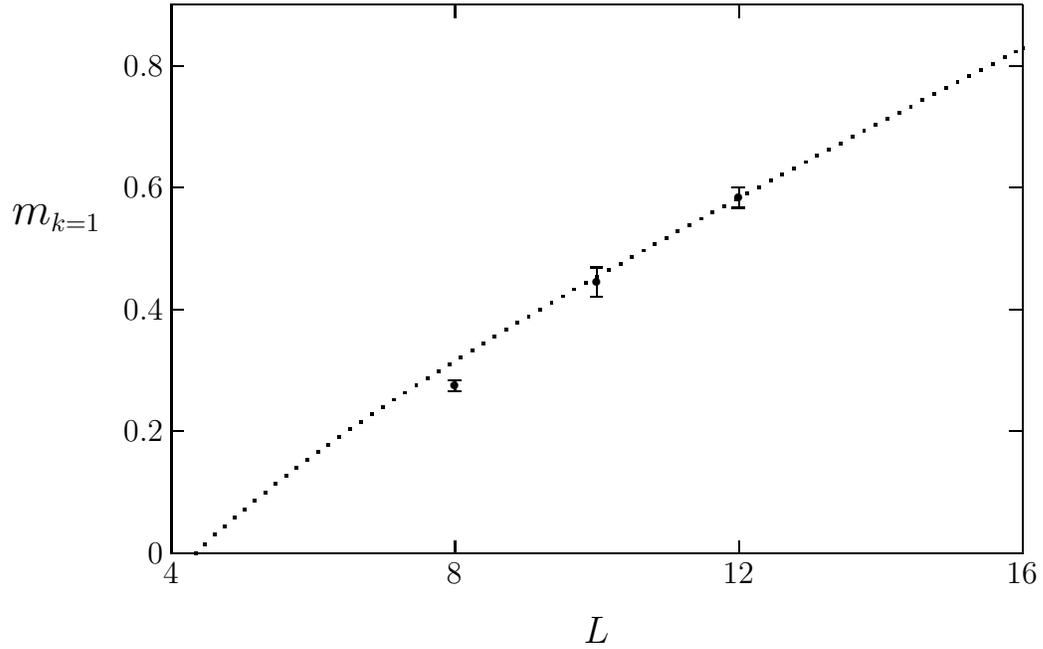

\end{document}